\documentclass[aps,prb,preprint,superscriptaddress,showpacs,amsmath]{revtex4}

\usepackage{graphicx,epsfig}

\begin{document}

\title{Exciton Gas Compression and Metallic Condensation in a Single Semiconductor Quantum Wire.}

\author{B. Al\'en}
\email[]{benito@imm.cnm.csic.es}

\author{D. Fuster}

\affiliation{IMM, Instituto de Microelectr\'onica de Madrid (CNM, CSIC), Isaac Newton 8,
 28760 Tres Cantos, Madrid, Spain.}

\author{G. Mu\~{n}oz-Matutano}

\author{J. Mart\'\i{}nez-Pastor}

\affiliation{ICMUV, Instituto de Ciencia de Materiales, Universidad de Valencia,
  P.O. Box 22085, 46071 Valencia, Spain.}

\author{Y. Gonz\'alez}

\affiliation{IMM, Instituto de Microelectr\'onica de Madrid (CNM, CSIC), Isaac Newton 8,
 28760 Tres Cantos, Madrid, Spain.}

\author{J. Canet-Ferrer}

\affiliation{ICMUV, Instituto de Ciencia de Materiales, Universidad de Valencia,
  P.O. Box 22085, 46071 Valencia, Spain.}

\author{L. Gonz\'alez}

\affiliation{IMM, Instituto de Microelectr\'onica de Madrid (CNM, CSIC), Isaac Newton 8,
 28760 Tres Cantos, Madrid, Spain.}

\date{\today}

\begin{abstract}
We study the metal-insulator transition in individual self-assembled quantum wires and report optical evidences of metallic liquid
condensation at low temperatures. Firstly, we observe that the temperature and power dependence of the single nanowire photoluminescence
follow the evolution expected for an electron-hole liquid in one dimension. Secondly, we find novel spectral features that suggest that in
this situation the expanding liquid condensate compresses the exciton gas in real space. Finally, we estimate the critical density and
critical temperature of the phase transition diagram at $n_c\sim1\times10^5$ cm$^{-1}$ and $T_c\sim35$ K, respectively.
\end{abstract}


\maketitle

The so-called metal-insulator transition in photoexcited semiconductor quantum wires (QWRs) has motivated many experimental and theoretical
investigations in the last
years.~\cite{Cingolani1991,Greus1996,Ambigapathy1997,Guillet2003,Yoshita2006,Ihara2007,Sarma2000,Wang2001,Asano2005,Huai2006} As known from
bulk semiconductors, the insulating exciton gas becomes unstable at high carrier densities and transforms into a many body conducting
state. Yet, in certain semiconductor systems, this transition occurs through condensation of the exciton gas into a metallic electron-hole
liquid (EHL), while in others, the exciton gas is completely ionized into a metallic electron-hole plasma (EHP). Despite their common
metallic character, both states have very different physical properties. An EHL, like any other liquid, is characterized by a well defined
interface with its surroundings and definite equilibrium density of e-h pairs.~\cite{Rice1977,Keldysh1986} Since the equilibrium density is
constant, an increment of the excitation increases the size of the crystal volume occupied by the liquid but does not change its internal
properties. On the contrary, an EHP is formed by unbound e-h pairs which fill the entire excited volume. In this case, a more intense
excitation leads to an increase of the EHP density and eventually to the disappearance of the excitons at the so-called Mott critical
density. The existence of 1D electron-hole plasmas in semiconductor QWRs has been evidenced in optical experiments performed both, in QWR
ensembles~\cite{Cingolani1991,Greus1996} and single QWRs~\cite{Guillet2003,Yoshita2006}. However, the observation of electron-hole liquids
in this kind of systems has been until know much more elusive. In this framework, in the following we shall present our results obtained in
individual InAs/InP self-assembled QWRs and demonstrate that the metal-insulator transition can occur in this system through compression of
the exciton gas and formation of a metallic liquid condensate, and not only through ionization of the exciton gas into a plasma as usually
reported.

\begin{figure}[htb]
\begin{center}
\includegraphics[width=58 mm]{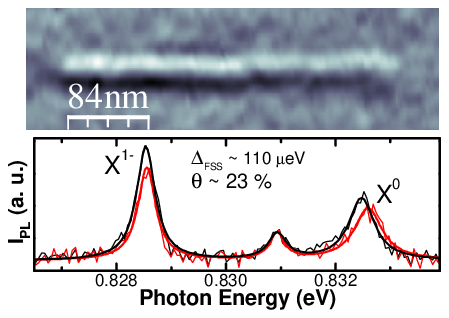}%
\caption{B. Al\'en et al.} \label{Fig1}
\end{center}
\end{figure}

The isolated InAs nanowires studied here have been obtained by self-assembling methods on InP(001) and present a large in-plane aspect
ratio with typical lengths exceeding more than ten times their average width $w\sim$~20 nm as shown in Fig.~\ref{Fig1}. More details about
the growth procedure, overall emission properties and confocal optical setup can be found elsewhere.~\cite{Fuster2007,Alen2006} At low
temperatures, their ground state emission energy at $\sim$~0.8 eV (1.55 $\mu$m) implies an estimated QWR height after capping of
$h\sim$~3.3 nm.~\cite{Alen2001} This small cross-sectional area, $w\times h$, leads to large subband energy spacings in both, the
conduction ($>30$ meV), and valence bands ($>10$ meV), and ensures that non-equilibrium carriers in these QWRs are effectively confined in
one dimension.

At 5.5 K and low excitation power, the single QWR photoluminescence spectrum shown in Fig.~\ref{Fig1} consists of several narrow peaks
whose linewidths, in the range of $\sim$0.3-0.5 meV, indicate a very good size uniformity.~\cite{Guillet2003,Yoshita2006} Due to their
elongated shape, the single QWR emission is slightly polarized along the QWR axis (1-10 crystal direction) yielding a polarization
anisotropy ratio $\theta=(P_0-P_{90})/(P_0+P_{90})\sim 23~\%$.~\cite{Alen2001} The asymmetric confinement also enhances the anisotropic
electron-hole exchange interaction (AEI) producing a relatively large polarization splitting of the neutral exciton line (X$^{0}$) at
0.8325 eV.~\cite{Hoegele2004} This splitting is absent when two electrons and one hole bond together into a negatively charged exciton or
trion (X$^{1-}$) as it can be observed for the main emission line at 0.8285 eV in Fig.~\ref{Fig1}. Narrow spectral features with similar
splittings and polarization behavior are systematically observed in our spectra at low excitation power. Weaker resonances, like the one
detected in this case at 0.8309 eV, appear occasionally and will be tentatively assigned to the emission from localized exciton states
whose anisotropy and splitting cannot be resolved due to their more symmetric wavefunctions.

\begin{figure}[htb]
\begin{center}
\includegraphics[width=73 mm]{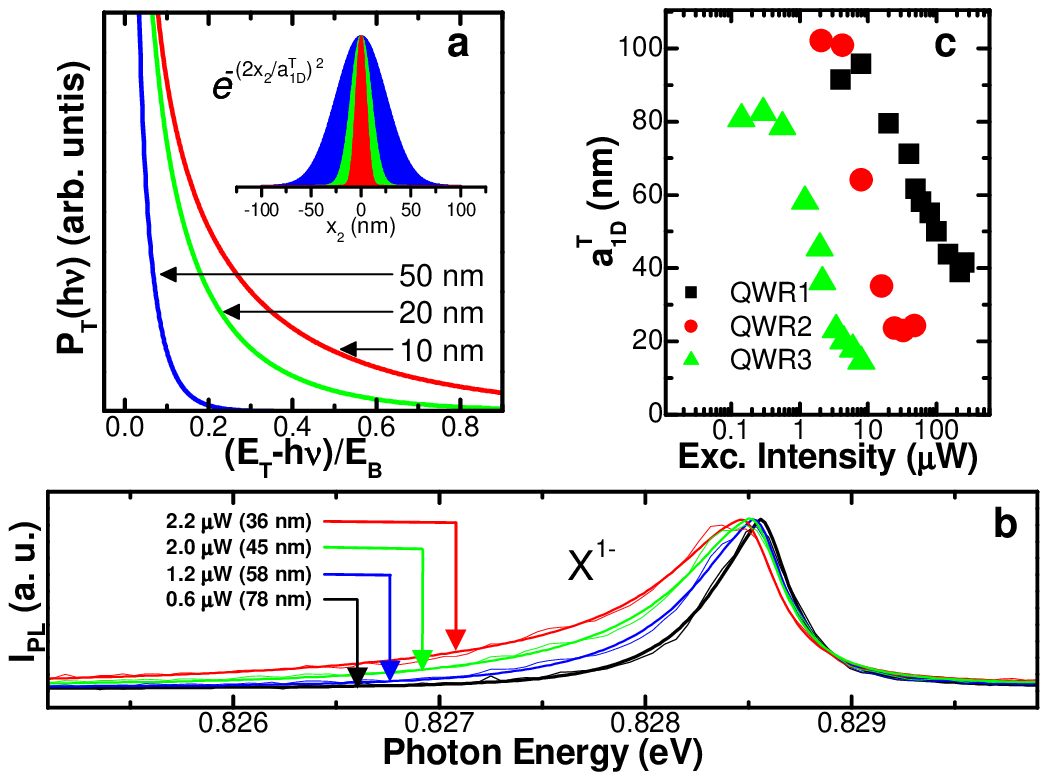}%
\caption{B. Al\'en et al.} \label{Fig2}
\end{center}
\end{figure}

The presence of negative trions in our spectra is not surprising given the rather high n-type residual doping of InP layers grown by
conventional molecular beam epitaxy methods. We have estimated $N_D\sim1\times10^{16}$ cm$^{-3}$ by Hall resistance measurements implying
an average occupation of at least two spectator electrons per QWR in absence of light. This fact is of central importance in our analysis
since the emission lineshape of the trion resonance strongly depends on its lateral extension and therefore on the size of the crystal
volume occupied by the exciton gas as explained below. This dependence can be established following the formulation based on earlier
theoretical work~\cite{Stebe1998} of A. Esser \textit{et al}~\cite{Esser2000} which, few years ago, studied the emission lineshape of
singly charged excitons in quantum wells (QWs). We have adapted here their results and considered a thermal distribution of 1D negative
trions with mass $M_T=2m_e+m_{hh}$ and energy $E_T+\hbar^2K_0^2/2M_T$. The emission lineshape is given by:~\cite{Esser2000}
\begin{equation}\label{Eq1}
P_T(\hbar\omega) \propto |M(K_0)|^2\textrm{exp}\left(-\frac{\epsilon}{k_BT}\frac{m_e}{M_X}\right)\frac{1}{\sqrt{\epsilon}}
\end{equation}
where $M_X=m_e+m_{hh}$ is the exciton mass, $\epsilon=E_T-\hbar\omega=\frac{\hbar^2K_0^2}{2m_e}\frac{M_X}{M_T}$ results from the energy
conservation, and the optical matrix element $M(K_0)$ is calculated separately from:
\begin{equation}\label{Eq2}
M(K_0)=\int_{-\infty}^{\infty} dx_2 \Psi_{X^{1-}}(x_1=0,x_2)\textrm{exp}\left(-i K_0 x_2 \frac{M_X}{M_T}\right)
\end{equation}
being $x_i$ the position of electron $i$ relative to the hole and $x_1=0$. Equation~\ref{Eq1} can be evaluated analytically for the special
case where the trion wavefunction, $\Psi_{X^{1-}}(0,x_2)$, is given by a gauss shape reducing Eq.~\ref{Eq2} to a simple Fourier transform
in momentum space. The resulting lineshapes have been depicted in Fig.~\ref{Fig2}(a) for three different gaussian lengths, $a^T_{1D}$, and
constant temperature, T=10 K. As it can be observed, the more localized is the trion in real space, the longer extends its
photoluminescence towards low energies, $\hbar\omega<E_T$. We have applied this scheme to our spectra by removing the inverse square root
singularity from Eq.~\ref{Eq1} in account of the finite trion lifetime ($\hbar/\Gamma_T$). Thick solid lines in Fig.~\ref{Fig2}(b)
represent the best fitting curves obtained varying $a^T_{1D}$ to describe the power evolution of the trion resonance identified before at
0.8285 eV. It must be noted that the lineshape evolution can be catched by the model meaning that the exciton trion wavefunction is being
apparently compressed in real space upon increase of the excitation intensity. To the best of our knowledge, such observation has not been
reported before in confined systems of any dimensionality.

The examination of the single QWR spectrum in a broader range can give us more information about the trion wavefunction shrinkage.
Increasing the laser power, a relatively broad band appears at the low energy side of the spectrum, as depicted in Figs.~3(a) and (b) for
two different single QWRs. At the same time, $a^T_{1D}$ diminishes from $\sim$100 nm to $\sim$20 nm as shown in Fig.~\ref{Fig2}(c). The
broad emission band lineshape evolution can be described using a 1D single band momentum conserving scheme with constant matrix elements
just as expected from the direct recombination of e-h pairs at the renormalized band gap edge.~\cite{Cingolani1991} The thick solid lines
in figures~3(a) and (b) are numerical fits obtained varying slightly the band gap energy, $E_g^i$, broadening parameter, $\Gamma_g$, and
carrier temperatures, $T_{e(h)}$, and keeping fixed the chemical potential energy, $\mu_{e}+\mu_{h}$, at the trion peak position. The
spectral analysis also allows the direct determination of the 1D carrier density, $n_{1D}$, from~\cite{Calleja1991}
$E_T-E_g^i=E_{fe}+E_{fh}=\frac{\pi^2\hbar^2 n_{1D}^2}{8m_e}(1+m_e/m_h)$ which can be used to discriminate between a condensed phase and a
plasma as explained below.

\begin{figure}[htb]
\begin{center}
\includegraphics[width=73 mm]{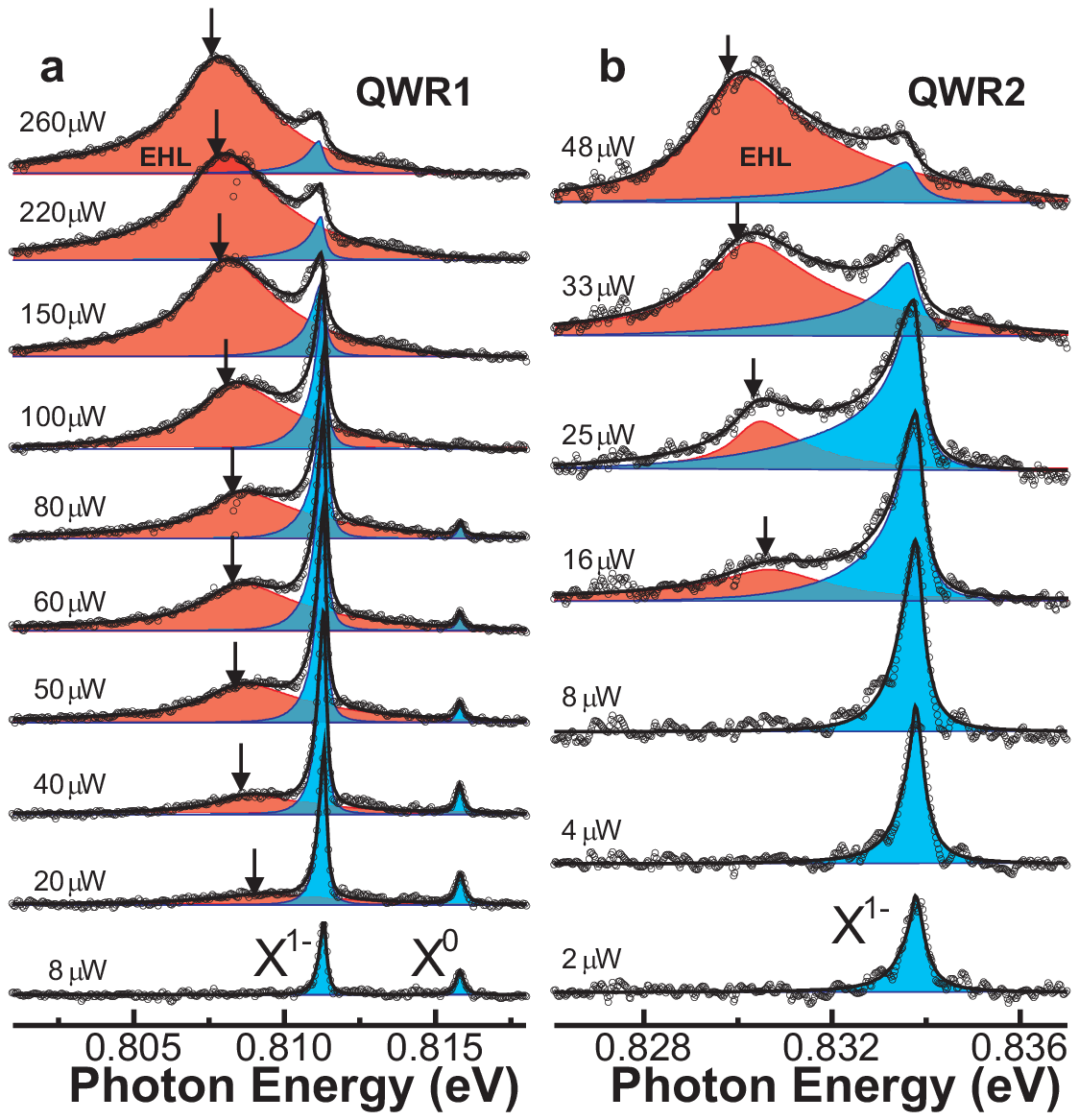}%
\caption{B. Al\'en et al.} \label{Fig3}
\end{center}
\end{figure}

The results depicted in Fig.~\ref{Fig3} represent a solid experimental evidence of the formation of an EHL in our system. At low excitation
powers, an e-h pair captured into the QWR can form a neutral exciton or, with the aid of an spectator electron, a negative trion. According
to our results, the binding energy of the negative complex is around 3 to 5 meV greater than the neutral one, and therefore trions will
prevail while excess electrons remain available.~\cite{Szafran2005} Above a certain excitation density, the overlap between individual
exciton complexes is such that many body excitations develop in the non-equilibrium carrier ensemble. At this point, several spectral
features reveal the presence of two spatially separated phases comprising an excitonic gas and an EHL. Firstly, the excitonic emission
peaks coexist with the band edge luminescence in a wide excitation window. Only when the latter is fully developed, the trion luminescence
gets appreciably quenched, as observed in Fig.~\ref{Fig3}. Secondly, as expected from a liquid phase with constant internal properties,
neither the position, nor the shape of the band edge emission changes significantly during the crossover. Indeed, despite the ten-fold
(QWR1) and three-fold (QWR2) increase of the excitation intensity, and hence of the number of carriers, $n_{1D}$ stays essentially
constant, and only varies between 2.5 and 2.9$\times10^{5}$ cm$^{-1}$, and between 2.7 and 3.0$\times10^{5}$ cm$^{-1}$, respectively. Our
results suggest a dynamical situation where after initial drop nucleation, the larger excitation increases the volume of the crystal
occupied by the liquid inducing the compression of the exciton gas against the QWR edges as discussed above.

\begin{figure}[htb]
\begin{center}
\includegraphics[width=73 mm]{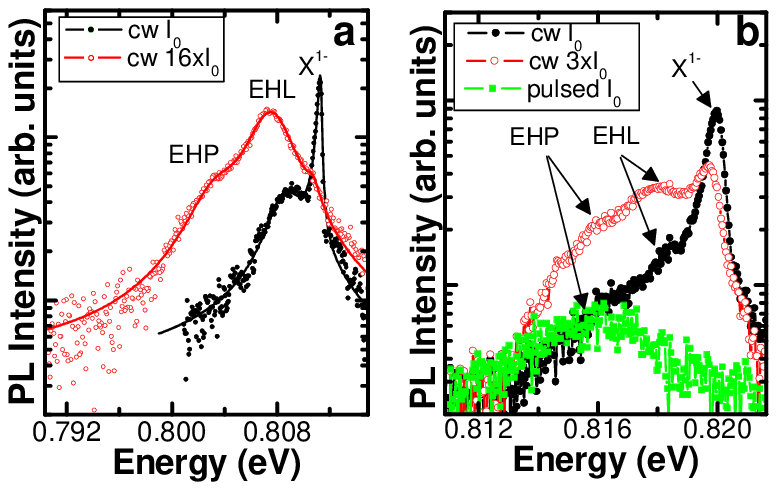}%
\caption{B. Al\'en et al.} \label{Fig4}
\end{center}
\end{figure}

If we increase the excitation power further, new spectral features appear in the tail of the EHL band. This is the case of the
semi-logarithmic plot of the spectrum of QWR1 excited with 800~$\mu$W and shown in Fig.~\ref{Fig4}(a). We tentatively assign the low energy
shoulder that appears at high excitation power to the formation of an electron hole plasma (EHP) in the QWR. The same occurs for the
nanowire represented in the adjacent figure where we used moderate excitation with a 980 nm 40 MHz pulsed laser (40 ps pulsed duration) to
deplete completely the excitonic and EHL emission and leave only the EHP band. We have observed the same behavior tuning the excitation
energy of our Ti:Saphire laser from 1000 nm to 865 nm. Only excitation well below the InP gap allowed the observation of sharp exciton
peaks and distinct EHL features in qualitative agreement with previous work done in highly excited direct gap
semiconductors.~\cite{Nagai2001}

\begin{figure}[htb]
\begin{center}
\includegraphics[width=80 mm]{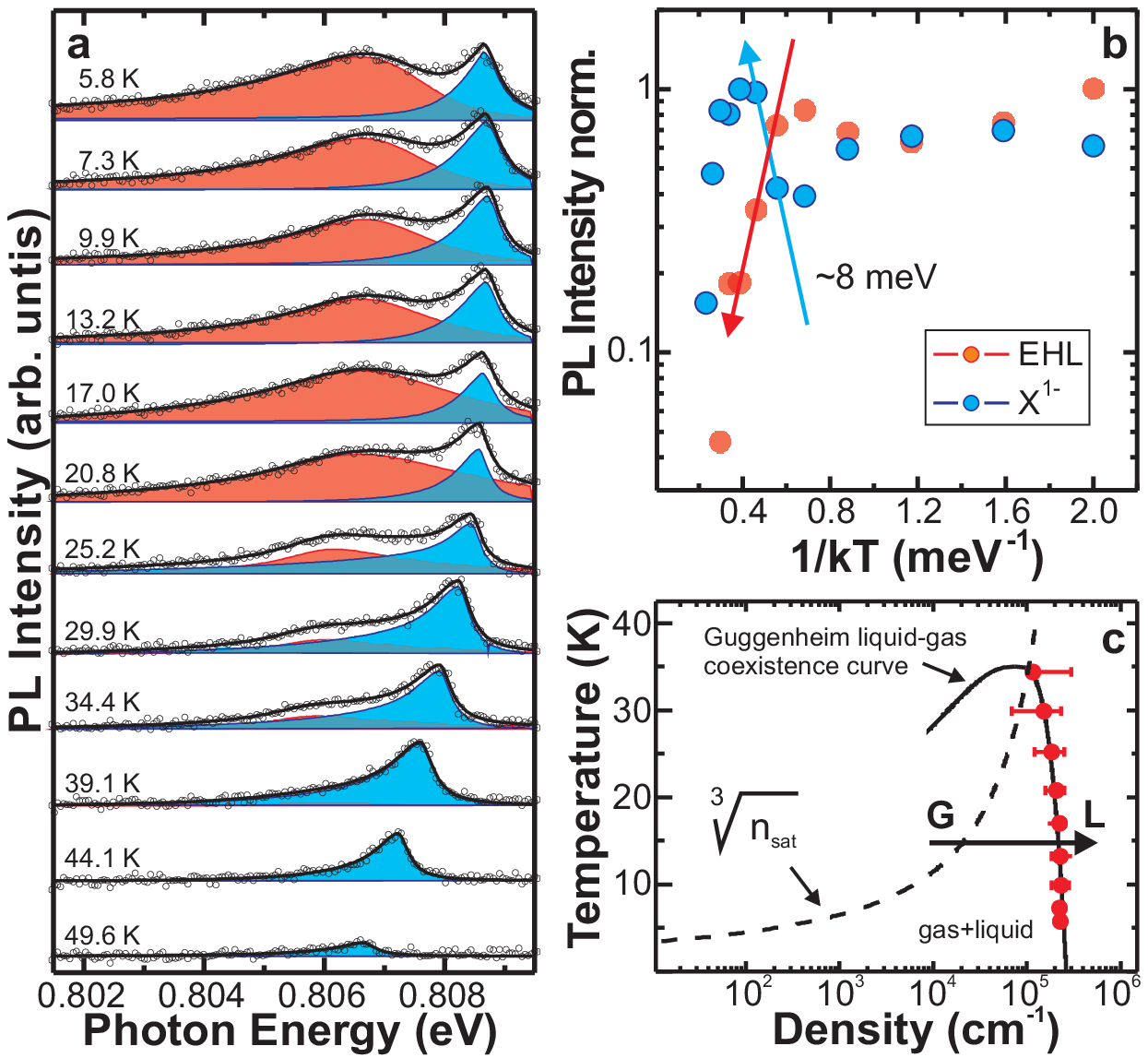}%
\caption{B. Al\'en et al.} \label{Fig5}
\end{center}
\end{figure}

The temperature dependence of the single QWR spectrum shown in figure~\ref{Fig5} corroborates our hypothesis. The presence of strong
quantum correlations within the many body ensemble is the main responsible of the appearance of a thermodynamically stable liquid phase
instead of an electron hole plasma in our system. Therefore, once the condensation has occurred, to extract an electron-hole pair out of
the condensed phase, some energy, $E_L$, must be provided.~\cite{Keldysh1986} This evaporation mechanism must be considered to explain the
evolution of the spectra shown in figure~\ref{Fig5}(a). Contrarily to the behavior expected for a metallic plasma, above $\sim$25 K the
liquid begins to evaporate and the system returns to its gaseous state. Furthermore, at the turning point, the reduction of the integrated
intensity of the EHL band and the enhancement of the trion emission peak evolve with the same rate. The value extracted from the slope of
the integrated intensity, $E_L\sim 8$ meV, fairly agrees with the energy splitting observed in most QWRs between the exciton peak ($X^0$)
energy and the one dimensional band edge position. Although the dynamic range of the fit is small, this could indicate that the evaporation
of the liquid is an ambipolar process followed by the trion formation at a later step.

Our results can be summarized in a typical phase diagram for a metallic EHL as shown in figure~\ref{Fig5}(c).~\cite{Keldysh1986} At low
temperatures ($T\leq$20 K) and moderate carrier densities ($n_{1D}\leq1\times10^4$ cm$^{-1}$), the system behaves as a diluted excitonic
gas (G) dominated by Coulomb interactions among only few particles. As the carrier density approaches $\sim10^{5}$ cm$^{-1}$, Coulomb
correlations become increasingly important and, in some regions of the excited volume, the exciton gas condensate into metallic
electron-hole droplets. The coexistence region (G+L) is limited by the equilibrium curves corresponding to the uniform gas (left branch)
and the uniform liquid (right branch). It comprises the range of parameter values where a spatially uniform distribution is not possible,
and there occurs a physical separation between well defined liquid and gaseous phases. The phase diagram shows a characteristic feature of
the electron-hole liquid: the density decreases when the temperature increases towards the critical point following a Guggenheim-like
liquid-gas coexistence curve.~\cite{Smith1995,Guggenheim1945} Therefore, the comparison of this curve with the data extracted from the
spectral lineshape analysis allows the determination, within our experimental uncertainty, of the critical temperature, $T_c$=35 K, and
critical density, $n_c=7.5\times10^4$ cm$^{-1}$, for this single quantum wire.

Finally, we must discuss briefly how our results could be affected by the QWR morphology and impurity environment. In around one out of
five studied QWRs, we have found that exciton localization induced by disorder of the confining potential prevents the formation of an EHL
phase. As reported for other QWR systems, in such cases, the strong localization leads to a high power $\mu$PL spectrum characterized by
radiative recombination of multiexciton complexes.~\cite{Guillet2003} Their behavior is very different of the disorder-free QWRs presented
in this work which show clear one-dimensional spectral features. Secondly, the impurity background has made possible the formation of
negatively charged excitons and therefore the observation of the exciton trion compressional effect. Since the random impurity potential
can pin the fermi level in different regions of the QWR,~\cite{Efros1989} weak localization of the electrons could make easier the initial
liquid drop nucleation in our system, although, at present, its precise role remains an open question.

In conclusion, we have identified a novel behavior studying the metal-insulator transition in individual semiconductor nanowires. Our
experimental results reveal that in this system this transition can occur through compression of the exciton gas and condensation in a
metallic liquid phase and should motivate future theoretical and experimental investigation of these and related excitonic and correlated
electron phenomena.

The authors acknowledge valuable discussions with Dr. Jose M. Llorens, Dr. Richard J. Warburton and Dr. Khaled Karrai. Financial support by
the Spanish MEC and CAM through projects No. TEC-2005-05781-C03-01/03, NAN2004-09109-C04-01/03, S-505/ESP/000200 and CSD2006-0019, and by
the European Commission through SANDIE Network of Excellence (No. NMP4-CT-2004-500101) is also gratefully acknowledged.


\newpage
\Large{\textbf{Figure Captions}}\\

\vspace{6 mm} \small

\textbf{Figure 1.-} Top panel: Atomic force micrograph of an uncapped sample showing an isolated self-assembled InAs/InP QWR. Bottom panel:
$\mu$PL spectra of a single QWR obtained at 5.5 K with linear polarizations selected along (black lines) and across (red lines) the QWR
axis.

\textbf{Figure 2.-} a) Evolution of the low energy emission tail calculated using Eq.\ref{Eq1} for charged excitons with different extent
in real space ($E_B$ stands for the trion binding energy). b) The increasing asymmetry of the experimental trion resonance (thin solid
lines) can be fitted to the model (thick solid lines). c) Excitation power evolution of the trion wavefunction length, $a^T_{1D}$, for
different QWRs.

\textbf{Figure 3.-} a) \& b) Emission spectra of different single QWRs upon increase of the excitation power. The lineshape analysis
reveals that the insulating excitonic gas (high energy peak) condensates into a metallic electron-hole liquid phase (low energy band).
Down-arrows stand for the position of the renormalized band edge in each spectrum.

\textbf{Figure 4.-} The panels show the emission spectra of two different QWRs obtained using the excitation conditions indicated in each
legend. In all cases the excitation wavelength was 980 nm and $I_0=50~\mu$W.

\textbf{Figure 5.-} a) Evolution of the single QWR emission spectrum with temperature. The excitation intensity was chosen at 5.8 K to make
visible both the excitonic and EHL contributions. b) Excitonic and band edge integrated intensities (normalized) as obtained from the
lineshape fits. c) Schematic phase diagram for the metal-insulator crossover in our system. The solid line represents the liquid-gas
coexistence curve which better describes our experimental data (solid circles). The dashed line shows the evolution of the saturated gas
density deduced roughly from the bulk InAs expression $n_{sat}=2\left(\frac{2\pi m^*_e k_B T}{h^2}\right)^{\frac{3}{2}}e^{\frac{-E_L}{k_B
T}}$ with $E_L=8$ meV.

\newpage

\newpage

\newpage

\newpage

\newpage

\end{document}